\documentclass[nofootinbib,floatfix,twocolumn,superscriptaddress,a4paper,prl]{revtex4}

\usepackage{amsmath}
\usepackage{amssymb}
\usepackage{graphicx}
\usepackage[T1]{fontenc} 
\usepackage{slashed}

\newcommand{\Br}{\text{Br}}

\newcommand{\AddrBonn}{%
Physikalisches Institut der Universit\"at Bonn, 53115 Bonn, Germany
}

\newcommand{\AddrVa}{%
 {\it AHEP Group, Instituto de F\'{\i}sica Corpuscular --
    C.S.I.C./Universitat de Val{\`e}ncia \\
    Edificio de Institutos de Paterna, Apartado 22085,
  E--46071 Val{\`e}ncia, Spain}
}

\newcommand{\AddrOrsay}{%
Laboratoire de Physique Th\'eorique, CNRS -- UMR 8627, 
Universit\'e de Paris-Sud 11\\ F-91405 Orsay Cedex, France}

\preprint{BONN-TH-2012-03,IFIC/12-09,LPT-12-12}
\begin{document}

\title{Enhancing $l_i \to 3 l_j$ with the $Z^0$-penguin}

\author{M. Hirsch} \email{mahirsch@ific.uv.es }
\affiliation{\AddrVa}

\author{F. Staub}\email{fnstaub@physik.uni-bonn.de}
\affiliation{\AddrBonn}

\author{A. Vicente} \email{avelino.vicente@th.u-psud.fr}
\affiliation{\AddrOrsay}

\begin{abstract}
Lepton flavor violation (LFV) has been observed in neutrino
oscillations.  For charged lepton FV decays only upper limits are
known, but sizable branching ratios are expected in many neutrino mass
models. High scale models, such as the classical supersymmetric
seesaw, usually predict that decays $l_i \to 3 l_j$ are roughly a
factor $\alpha$ maller than the corresponding decays $l_i \to l_j
\gamma$. Here we demonstrate that the $Z^0$-penguin diagram can give
an enhancement for decays $l_i \to 3 l_j$ in many extensions of the
Minimal Supersymmetric Standard Model (MSSM). We first discuss why the
$Z^0$-penguin is not dominant in the MSSM with seesaw and show that
much larger contributions from the $Z^0$-penguin are expected in
general. We then demonstrate the effect numerically in two example
models, namely, the supersymmetric inverse seesaw and R-parity
violating supersymmetry.
\end{abstract}
\maketitle

{\em Introduction:}
Neutrino oscillation experiments \cite{Fukuda:1998mi,Ahmad:2002jz} 
have firmly established that lepton flavor is violated in the neutrino 
sector, with two of the three measurable mixing angles being surprisingly 
large. Observation of the characteristic ``neutrino dip'' 
leaves no doubt that neutrinos have mass \cite{Ashie:2004mr,arXiv:0801.4589} 
and quite accurate values for the mass squared differences are known 
now
\cite{Schwetz:2011zk}.  
In the charged lepton sector, however, only upper limits on LFV 
branching ratios, such as $\mu \to e \gamma$ \cite{Adam:2011ch} 
or $\mu \to 3 e$ \cite{Nakamura:2010zzi}, exist. 

Extending the standard model (SM) only by neutrino masses does not 
automatically lead to measurable charged LFV (CLFV), but sizable 
branching ratios are expected in many models. In fact, on quite 
general grounds one expects large CLFV, if physics beyond the SM 
exists at the TeV scale. A prime example for this observation is 
supersymmetry (SUSY). Here, the mass matrices of the new scalar particles 
need not (and in general will not) be aligned with those of the SM 
fermions. CLFV decays will occur and one can estimate roughly the 
branching ratios for radiative decays as \cite{Kuno:1999jp}
\begin{equation}
 \Br(l_i \to l_j \gamma) \simeq \frac{48 \pi^3 \alpha}{G_F^2}
\frac{|(m^2_{\tilde f})_{ij}|^2}{m^8_{SUSY}} \Br(l_i \to l_j \nu_i{\bar\nu_j})
\end{equation}
where $(m^2_{\tilde f})_{ij}$ parameterizes the dominant off-diagonal
elements of the soft SUSY breaking slepton mass matrices and $m_{SUSY}$ 
is the typical mass of the SUSY particles, expected to be in the 
ballpark of ${\cal O}(0.1-1)$ TeV. Rather small off-diagonal elements are 
required to satisfy experimental bounds \cite{Adam:2011ch,Nakamura:2010zzi}.

In the Constrained Minimal Supersymmetric extension of the SM (CMSSM), 
on the other hand, CLFV is zero, just as in the SM, simply because 
neutrinos are assumed to be massless. Extending the CMSSM to include 
neutrino masses (and mixings), for example by a seesaw mechanism, then 
leads to CLFV decays, because the flavor violation necessarily present 
in the Yukawa couplings is transmitted to the slepton mass matrices 
in the RGE (renormalization group equation) running \cite{Borzumati:1986qx}. 
In such high-scale neutrino mass models, with only MSSM particle content 
at the electroweak scale, it has been shown that the photonic penguin 
diagram gives the dominant contribution to $l_i \to 3 l_j$ decays in 
large regions of parameter space.\footnote{An exception from this
rule is the decay $\tau \to 3 \mu$ in the limit of large $\tan\beta$
\cite{Babu:2002et}.} In this case a simple relation 
can be derived \cite{Arganda:2005ji}
\begin{equation}
\Br(l_i \to 3 l_j) \simeq \frac{\alpha}{3 \pi} 
\left(\log\left(\frac{m^2_{l_i}}{m^2_{l_j}}\right) - \frac{11}{4} \right) 
\Br(l \to l' \gamma)
\end{equation}
Thus, usually it is concluded that the decays $l_i \to l_j \gamma$ 
are more constraining than the decays $l_i \to 3 l_j$.

Apart from the photonic penguin, there are also box diagrams, Higgs- and  
$Z^0$-penguin contributing to the decays $l_i \to 3 l_j$. The latter 
diagram is not {\em per se} smaller than the photonic penguin. Rather, 
as we will show below, in models with only the MSSM particle content 
and couplings, the $Z^0$-penguin is suppressed by a subtle cancellation 
among different terms in the amplitude. Such a cancellation, however, 
can be easily spoiled if there are (a) new couplings and/or (b) a 
larger particle content than in the MSSM. Then, as we will discuss, 
the $Z^0$-penguin can easily give the dominant contribution to 
$l_i \to 3 l_j$. We will demonstrate this fact numerically with two 
typical example models: (i) a supersymmetric inverse seesaw and 
(ii) R-parity violating SUSY. The former is an example of 
a model with extended particle content, while the latter is an example 
of a model with the MSSM particle content but new interactions. As we 
will show, in such models $l_i \to 3 l_j$ can be more constraining 
than $l_i \to l_j \gamma$. This is the main result of the present paper.

Finally, we emphasize that the $Z^0$-penguin can be dominant also in
other observables and for other theoretical models, although this fact
has not, in general, been discussed before. For example,
$Z^0$-dominance can be found in $\mu-e$ conversion in nuclei in
supersymmetric models with R-parity violation, as can be seen from the
numerical results of reference \cite{Faessler:2000pn}, although the
authors do not discuss it. Similarly, in the little Higgs model of
\cite{Goto:2010sn} one finds numerical results with parameter points
where $\Br(l_i \to 3 l_j) > \Br(l_i \to l_j \gamma)$, despite the
authors concluding that both are correlated.

{\em Analytical discussion:} 
The total width of the $l_i \to 3 l_j$ decay contains contributions from 
the photon penguin, the Higgs penguin, the $Z^0$-penguin and boxes. 
Considering only contributions from photon and $Z^0$-penguin, which are 
the ones of interest to us and numerically the most important ones, the 
total width $\Gamma \equiv \Gamma(l_i^- \to l_j^- l_j^- l_j^+)$ can be 
written as~\cite{Hisano:1995cp}:
\begin{widetext}
\begin{eqnarray}
\Gamma &=& \frac{e^4}{512 \pi^3} m_{l_i}^5 
\Big[ \left| A_1^L \right|^2 + \left| A_1^R \right|^2 
- 2 \left( A_1^L A_2^{R \ast} + A_2^L A_1^{R \ast} + h.c. \right)
+ \left( \left| A_2^L \right|^2 + \left| A_2^R \right|^2 \right) 
\left( \frac{16}{3} \log{\frac{m_{l_j}}{m_{l_i}}} - \frac{22}{3} \right) 
\nonumber \\
&+& \frac{1}{3} 
\left\{ 2 \left( \left| F_{LL} \right|^2 + \left| F_{RR} \right|^2 \right) 
+ \left| F_{LR} \right|^2 + \left| F_{RL} \right|^2\right\} + I_{AF} \Big]
\label{decay}
\end{eqnarray}
\end{widetext}
Here, terms denoted $A$ ($F$) are due to photon ($Z^0$) exchange and 
$I_{AF}$ denotes their interference terms, irrelevant for the following 
discussion. Both, photon and $Z^0$ penguins have chargino-sneutrino 
and neutralino-slepton contributions. Exact definitions can be found 
in \cite{Arganda:2005ji}. We will focus on the chargino loops 
for brevity here, since the effects we are interested in are most 
pronounced in these loops. The photon contributions are 
\begin{equation} \label{Achar}
A_a^{(c)L,R} = \frac{1}{m_{\tilde{\nu}}^2} {\cal O}_{A_a}^{L,R}s(x^2)
\end{equation}
whereas the $Z$-contributions read
\begin{equation} \label{Fchar}
F_{X} = \frac{1}{g^2 \sin^2 \theta_W m_Z^2}{\cal O}_{F_X}^{L,R}t(x^2)
\end{equation}
with $X=\left\{LL,LR,RL,RR\right\}$. In these expressions 
${\cal O}_{y}^{L,R}$ denote combinations of rotation matrices 
and coupling constants and $s(x^2)$ and $t(x^2)$ are short-hands 
for the Passarino-Veltman loop functions which depend on $x^2 =
m_{\tilde{\chi}^-}^2/m_{\tilde{\nu}}^2$. For precise definitions 
see  \cite{Arganda:2005ji}.

The scaling $A \sim m_{SUSY}^{-2}$ and $F \sim m_Z^{-2}$ can 
be understood, in principle, from simple dimensional analysis. 
The width of the decay is proportional to $m_{l_i}^5$, so both 
$A$ and $F$ must be $A,F \propto m^{-2}$. In this case
it is the smallest mass term in the loop which sets the scale, 
which in $F$ is $m_Z$. Due to the masslessness of the photon 
in case of $A$ the smallest mass scale in the loop is $ m_{SUSY}$. 
With $m_Z^{2} \ll m_{SUSY}^2$ the $Z^0$ penguin can, in 
principle, be even more important than the photonic one.

Numerically, however, it has been found in case of the MSSM 
that the photonic penguin is dominant \cite{Arganda:2005ji}. 
This can be understood as follows. To simplify the discussion, 
we neglect first $F_R$, since it is always proportional to the 
charged lepton Yukawa couplings. Consider then $n$ generations of 
sneutrinos and neglect the chargino mixing. In this simplified scenario 
only the wino contributes to $F^{(c)}_L$ and it can be written as
\begin{equation}\label{sum-M-FL}
F^{(c)}_L = M_{\text{wave}} + M_{\text{p1}} + M_{\text{p2}} 
\end{equation}
with 
\begin{eqnarray}
M_{\text{wave}} &=& \frac{1}{2} g^2 (g c_W - g' s_W) 
                    Z_V^{ik} Z_V^{ij*} f_{\text{wave}}^i \\
M_{\text{p1}} &=& - g^3 c_W Z_V^{ik} Z_V^{ij*} f_{\text{p1}}^i \\
M_{\text{p2}} &=& \frac{1}{2} g^2 (g c_W + g' s_W) 
                   Z_V^{ik} Z_V^{ij*} f_{\text{p2}}^i
\end{eqnarray}
Summing over the index $i$ is implied. The terms in the sum come
from different types of diagrams: wave function diagrams
($M_{\text{wave}}$), penguins with the $Z^0$-boson attached to the
chargino line ($M_{\text{p1}}$) or the sneutrino line ($M_{\text{p2}}$). 
Moreover, $c_W = \cos
\theta_W$, $s_W = \sin \theta_W$, $Z_V$ is a $n \times n$ unitary
matrix that diagonalizes the mass matrix of the sneutrinos and we used
the abbreviations $f_{\text{wave}}^i = -
B_1(m_{\tilde{\chi}^\pm}^2,m_{\tilde{\nu}_i}^2), f_{\text{p1}}^i =
\frac{1}{2}
\tilde{C}_0(m_{\tilde{\nu}_i}^2,m_{\tilde{\chi}^\pm}^2,m_{\tilde{\chi}^\pm}^2)
- m_{\tilde{\chi}^\pm}^2
C_0(m_{\tilde{\nu}_i}^2,m_{\tilde{\chi}^\pm}^2,m_{\tilde{\chi}^\pm}^2),
f_{\text{p2}}^i = \frac{1}{2}
\tilde{C}_0(m_{\tilde{\chi}^\pm}^2,m_{\tilde{\nu}_i}^2,m_{\tilde{\nu}_i}^2)$. 
The sum in eq.~\eqref{sum-M-FL} vanishes exactly as can
be seen by grouping the different terms 
\begin{equation}
F^{(c)}_L = \frac{1}{2} g^3 c_W Z_V^{ik} Z_V^{ij*} X_1^i 
+ \frac{1}{2} g^2 g' s_W Z_V^{ik} Z_V^{ij*} X_2^i
\end{equation}
with $X_1^i = f_{\text{wave}}^i - 2 f_{\text{p1}}^i + f_{\text{p2}}^i,
X_2^i = f_{\text{p2}}^i - f_{\text{wave}}^i$. Using the exact
expressions for the loop functions \cite{Arganda:2005ji} one finds
that the masses cancel out and these combinations become just
numerical constants: $X_1^i= - \frac{3}{4}$ and $X_2^i = -
\frac{1}{4}$. Therefore, one is left with $F^{(c)}_L \propto \sum_i
Z_V^{ik} Z_V^{ij*}$, which vanishes due to unitarity of the $Z_V$
matrix\footnote{In reference \cite{Lunghi:1999uk}, where the authors
  study $B \to X_s l^+ l^-$ in supersymmetry, the Z-penguin
  contributions are found to be sub-dominant due to the same type of
  cancellation that is found in our work.}.

This cancellation can be spoiled by two effects, either (i) the
sneutrinos mix with other particles which are not $SU(2)_L$ doublets
so that the factorization no longer holds, or (ii) the charginos
are not pure wino and higgsino states. The last effect is of course
present in the MSSM and therefore this cancellation is not
exact. Nevertheless, the $Z^0$-contributions are suppressed due to their
proportionality to the square of the chargino mixing angle (two
wino-higgsino insertions are necessary since there is no
$\tilde{H}^\pm-\tilde{\nu}_L-l_L$ coupling). 
We neglected so far Higgsino interactions because in many 
models these couplings are very small in comparison to the 
gauge interactions (for example, a SUSY scale type-I seesaw model would 
have $Y_\nu \sim 10^{-6}$). 
However, in models where the Higgsino can have much larger
Yukawa interactions, a large enhancement of the $Z^0$-contributions can 
be expected. This will be addressed numerically in the next section.

\begin{figure}[hbt]
\includegraphics[width=0.9\linewidth]{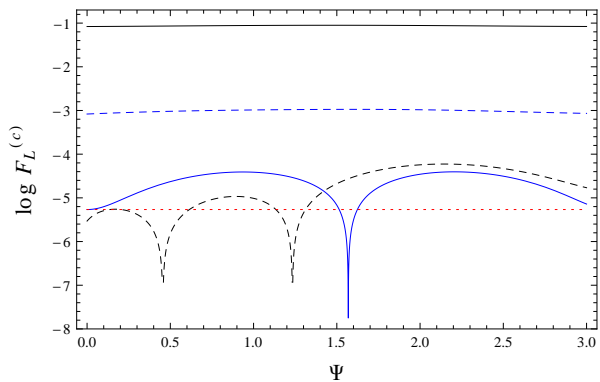}
\caption{$F^{(c)}_L$ for our toy model as a function of the
sneutrino left-right mixing angle $\Psi$ and for different values of
$\kappa$: $10^{-4}$ (blue), $10^{-2}$ (black dashed), 0.1 (blue dashed) and 1.0
(black).}
\label{fig:approx}
\end{figure}

Before turning to a numerical discussion, we consider for simplicity 
a toy model consisting of two
generations of left-handed sneutrinos which can mix with one
generation of right-handed sneutrinos. A $3\times 3$ rotation matrix
is in general parametrized by 3 angles, but we will assume here for
simplification that two of them vanish and call the third one
$\Psi$. In addition, we introduce a new interaction for the Higgsinos
$\kappa \nu^c \tilde{H}_u \tilde{l}_L$. We give in
Fig.~\ref{fig:approx} the computed $F_L^{(c)}$ for arbitrarily chosen
sneutrino and chargino masses as a function of $\Psi$ for different
values of $\kappa$. The red dotted line shows the case for
$\Psi=\kappa=0$. As clearly seen, $F_L^{(c)}$ depends on the
left-right mixing already for small values of $\kappa$. However, 
increasing $\kappa$, $F_L^{(c)}$ becomes totally dominated by
the new $\kappa$ interactions and enhances $\text{Br}(l_i \to 3
l_j)$.

{\em Numerical examples:} We turn to the full fledged numerical study of 
two examples: an inverse seesaw model and the MSSM with
$R$-parity violation. For this purpose, we have created for both
models {\tt SPheno} modules \cite{Porod:2003um,Porod:2011nf} using the
Mathematica package {\tt SARAH}
\cite{Staub:2011dp,Staub:2010jh,Staub:2009bi}. These
modules calculate the low-energy observables exactly including 
all possible diagrams \cite{Staub:2011dp}.

{\em Inverse Seesaw:} 
In inverse seesaw the MSSM particle content is extended by
three generations of right-handed neutrino superfields $\hat{\nu}^c$
and of gauge singlets $\hat{N}_S$ which carry lepton number
\cite{Mohapatra:1986bd,DeRomeri:2011ie}. The superpotential reads
\begin{equation}
\label{eq:W_lin_inv}
 W_{IS} = W_{\text{MSSM}} + Y_{\nu}\hat{\nu}^c\hat{L}\hat{H}_u
+M_R\,\hat{\nu}^c{\hat{N}}_{S}
+ \frac{\mu_N}{2}{\hat{N}}_{S}{\hat{N}}_{S} \\
\end{equation}
After electroweak symmetry breaking (EWSB) the effective mass matrix 
for the light neutrinos is approximately $m_\nu \simeq \frac{v_u^2}{2}
Y_\nu (M^T_R)^{-1} \mu_N M^{-1} Y^T_\nu$. Since $\mu_N$ can be of
${\cal O}(10^{-1})$ keV or even smaller while $M_R$ is of 
${\cal O}(m_{SUSY})$, the neutrino Yukawa couplings have to be much 
larger than for a standard weak-scale seesaw to explain neutrino data. 

\begin{figure}[bt]
\includegraphics[width=0.9\linewidth]{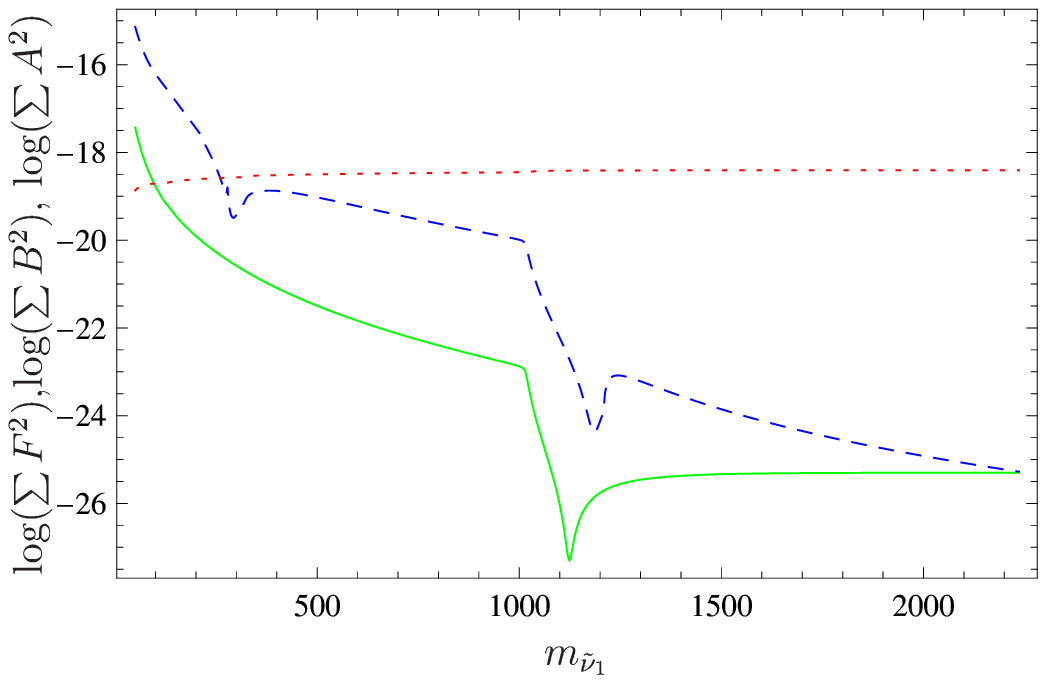} \\
\includegraphics[width=0.9\linewidth]{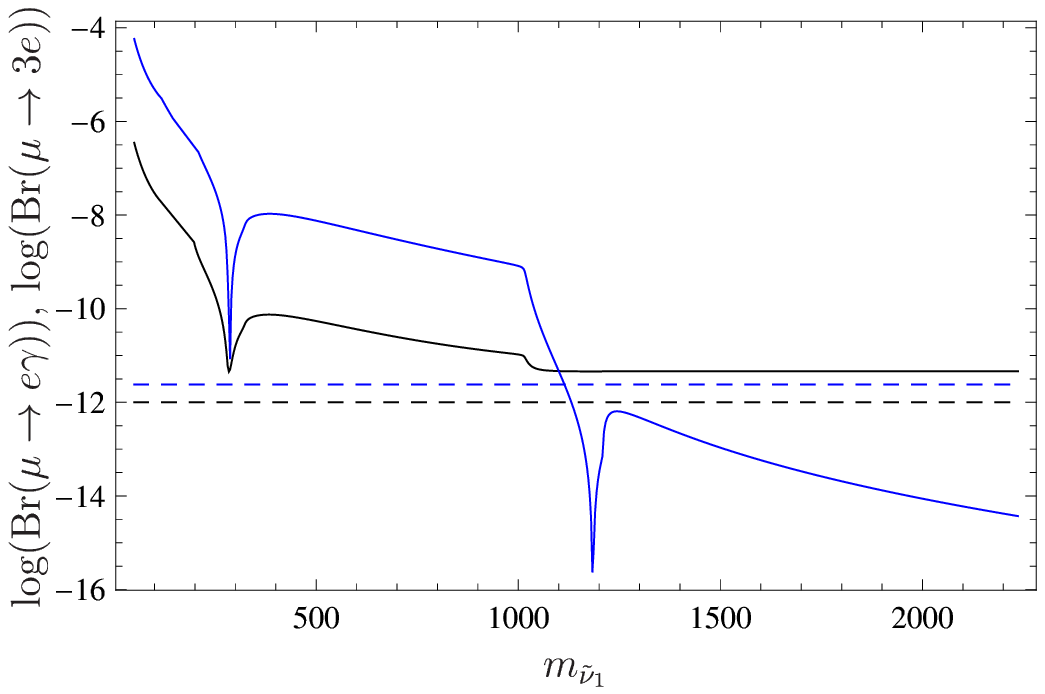}
\caption{Top: Different contributions to $\Br(\mu \to 3 e)$ as
function of the lightest sneutrino mass: $Z^0$-penguins (red dotted), photonic
penguins (blue dashed), combined Higgs-penguins/box diagrams
(green). Bottom: $\Br(\mu \to 3 e)$ (black) and $\Br(\mu \to e
\gamma)$ (blue) and the current experimental bounds (dashed
lines). The dips are an effect of a mass crossing between charginos
and sneutrinos. }
\label{fig:InvSeesaw}
\end{figure}

Due to the extended particle content, new contributions for 
$\text{Br}(l_i \to 3 l_j)$ are expected in the inverse seesaw. 
For example, the Higgs mediated contributions were recently studied 
in \cite{Abada:2011hm}. In Fig.~\ref{fig:InvSeesaw} (top) we show the 
different contributions to $\Br(\mu \to 3 e)$ for a variation of the 
SUSY masses. To disentangle RGE effects we have calculated once the 
spectrum for a CMSSM input ($m_0 = 500$~GeV, $M_{1/2} = 1$~TeV, 
$\tan(\beta) = 10$, $A_0=-300$~GeV) and rescaled all dimensionful 
parameters at the SUSY scale. This changes the sfermion masses but not 
the mixing matrices.  $Y_\nu$ has been chosen to explain neutrino 
data for $\text{diag}(\mu_N) = 10^{-1}$ keV and $M_R=1$~TeV. Clearly, 
the $Z^0$-penguins dominate and are nearly independent of the SUSY 
scale. Only in the limit $m_{SUSY} \to m_Z$ the other contributions 
can compete. In Fig.~\ref{fig:InvSeesaw} (bottom) the branching ratios 
for $\mu \to e \gamma$ and $\mu \to 3 e$ and the current 
experimental bounds of $2.4\cdot 10^{-12}$ and $1.0\cdot 10^{-12}$ are 
depicted \cite{Adam:2011ch,Nakamura:2010zzi}. While $\Br(\mu \to e \gamma)$ 
would be in conflict with experiment only for $m_{\tilde\nu_1} < 1.2$~TeV, 
$\Br(\mu \to 3 e)$ rules out the entire range.
In this example, we have assumed $\mu_N$ and $M_R$ to be diagonal 
and all flavor violation comes from $Y_\nu$, as is usually done in
literature. However, neutrino oscillation data could equally well be 
fitted with the flavor violation coming from $\mu_N$ and $M_R$. 
In that case CLFV observables would be much smaller and
consistent with experimental data. However, the relative order between
2- and 3-body decays won't change, i.e.  $\Br(l_i \to 3 l_j)$ will
be most likely observed before $\Br(l_i \to l_j \gamma)$ if inverse
seesaw is realized in nature.

\begin{figure}[t]
\includegraphics[width=0.9\linewidth]{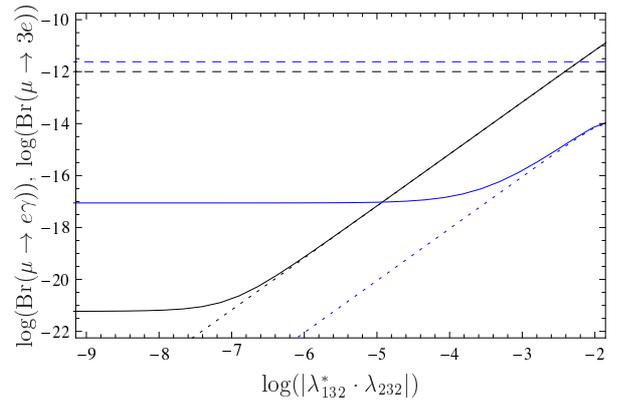} \\
\caption{$\Br(\mu \to 3 e)$ (black) and $\Br(\mu \to e \gamma)$ (blue)
varying $|\lambda_{132}^* \cdot \lambda_{232}|$ in  a mixed bi- and trilinear  
(solid lines) 
and a pure tri-linear (dotted lines) $R$pV scenario. The dashed lines show
the experimental limits.}
\label{fig:TRpV}
\end{figure}

{\em $R$-parity violation:} As second example, we take the MSSM
particle content but extend the superpotential by the lepton number
violating terms \cite{Allanach:2003eb}
\begin{equation}
 W_{\slashed{R}} = W_{MSSM} 
 + \frac{1}{2}  \lambda_{ijk} \hat{L}_i \hat{L}_j \hat{E}^c_k 
 + \lambda^{'}_{ijk} \hat{L}_i \hat{Q}_j \hat{D}^c_k 
 + \epsilon_i \hat{L}_i \hat{H}_u 
\end{equation}
While the $\epsilon$-parameters are highly constrained by neutrino
data \cite{Hirsch:2000ef}, the bounds for the tri-linear couplings are
much weaker and some entries can be of ${\cal O}(1)$
\cite{Dreiner:2010ye}. In the following, all entries of $\lambda$ and
$\lambda^{'}$ are set to zero but $\lambda_{132}$ and $\lambda_{232}$.
We give in Fig.~\ref{fig:TRpV} the results for $\Br(\mu \to 3 e)$ and
$\Br(\mu \to e \gamma)$ for a mixed bi- and tri-linear as well as for
the pure tri-linear scenario varying $|\lambda^*_{132}\cdot
\lambda_{232}|$. In the mixed case $\epsilon_i$ and the vacuum
expectation values of the sneutrinos, $v_L^i$, have been chosen to be
consistent with neutrino data and a moderate flavor violation in the
sneutrino sector has been induced by $m_{\tilde{l}_iH_d}^2 = (45
\hskip1mm {\rm GeV})^2$. It can be seen that in the mixed case
$\Br(\mu \to 3 e) > \Br(\mu \to e \gamma)$ holds when
$|\lambda_{132}^* \cdot \lambda_{232}|$ crosses $2.5 \cdot 10^{-5}$, while
for the pure tri-linear case without any flavor violation at tree
level in the sneutrino sector the three body decays dominate even for
much smaller values.

In both cases we get an upper limit for $|\lambda_{132}^* \cdot
\lambda_{232}|$ of $2.5 \cdot 10^{-3}$ from the bounds on $\Br(\mu \to
3e)$ for sneutrino masses of 730~GeV. So far, in the literature just 
the limits for $m_{\tilde{\nu}} = 100$~GeV from the photonic penguins 
\cite{Barbier:2004ez} have been published. These are much weaker,  
after rescaling the bound $\sim 7.1\cdot 10^{-5}
\left(\frac{730~\text{GeV}}{100~\text{GeV}}\right)^4 \simeq 0.2$.

{\em Summary:} 
We have shown in this letter that the $Z^0$-penguin can give the 
dominant contribution in lepton flavor violating three body decays 
in many models. The importance of the $Z^0$-penguin increases with 
increasing SUSY particles masses. As numerical examples, we have 
briefly discussed the supersymmetric inverse seesaw and the MSSM 
with $R$-parity violation.

\section*{Acknowledgements}

We thank Werner Porod, Debottam Das and Daniel E. Lopez-Fogliani for 
fruitful discussions. 
A.V. acknowledges support from the ANR project CPV-LFV-LHC {NT09-508531}.
This work was supported by the Spanish MICINN under grants
FPA2008-00319/FPA and FPA2011-22975 by the MULTIDARK Consolider 
CSD2009-00064, by the Generalitat Valenciana grant Prometeo/2009/091 
and by the EU grant UNILHC PITN-GA-2009-237920.

\end{document}